\documentstyle[11pt,paspconf,epsf]{article}

\begin{document}

\title{Dark Matter in Disc Galaxies}
\author{A. Bosma}
\affil{Observatoire de Marseille, 2 Place Le Verrier, 13248 Marseille
Cedex 4, France}

\begin{abstract}
Recent work on the mass distribution in spiral galaxies, using 
mainly HI observations, is reviewed. The principal problem is 
still to determine to what extent the dark matter is important in 
the inner parts of a galaxy, or in other words, how dominant is the 
self-gravitation of the disc. Studies of the shapes of rotation
curves show that in detail there is sufficient individuality in 
spiral galaxies to prohibit the construction of ``Universal Rotation
Curves''. A detailed account is given of the method of Athanassoula 
et al. (1987), where swing amplifier criteria are applied to set a 
range in the mass-to-light ratio of the disc. To restrict this range 
further, other methods might be useful. For a number of bright spirals
the rotation curve drops just outside the optical image, but this 
feature by itself cannot constrain unambiguously the mass models. 
The use of velocity dispersions seems a promising way, though the 
observational problems are hard. Within the uncertainties, discs 
can be close to ``maximum'', even though a range of values cannot 
be excluded.
\end{abstract}

\keywords{Spiral galaxies, Dark matter}

\section{Introduction}

The standard way to construct a mass model for a spiral galaxy is
to make composite disc/halo models assuming a ``maximum disc''
solution, or to adopt a ``best fit''. In such models, the data from 
surface photometry are used, assuming constant mass-to-light ratios, 
to calculate the expected rotation curve for the visible components, 
bulge and disc. From the observed HI gas density, and a suitable
factor to include helium, a rotation curve is calculated also, and 
quadratically added to the first one. The resulting curve is then 
compared with the observed rotation curve, and an additional dark 
halo component is introduced when necessary.

For extended HI rotation curves, such an analysis has been done by 
several authors, e.g. Begeman, Broeils \& Sanders (1991). The
constancy of the mass-to-light with radius within each component is 
usually justified by the absence of colour gradients. Indeed, data of 
De Jong (1996) shows that colour gradients are small, and if present, 
they are in the sense that discs become bluer outwards. In that case
the use of near infrared data is preferred, since it accounts better
for the contribution to the mass of the old stellar disc. From De
Jong's (1996) data, it follows that the ratio of the blue (B-band)
and near infrared (K'-band) scalelengths is about 1.15 for 85\% of his 
sample. However, from Verheijen's studies on the spirals in the Ursa 
Major cluster (Verheijen 1997), a much larger spread in the ratio
of scalelengths is seen (cf. Tully et al. 1996).

The primary trend in these models is that the importance of dark
matter increases with decreasing luminosity : small galaxies seem
to be more dark matter dominated than large galaxies (cf. Broeils
1992). Another trend is that the rotation curve of the gas component
has more or less the same shape as the halo rotation curve : if all
matter were is a disc, the ratio of gas mass to total mass becomes
constant in the outer parts (cf. Bosma 1978, 1981b; Carignan et al.
1990).

However, such results, based as they are on the assumption of maximum
disc, may not be correct. In this review, I will comment on the
recent work on rotation curves and mass models of spiral galaxies,
and in particular address part of the question whether the maximum 
disc hypothesis is justified. For another discussion on this last
topic, using arguments principally concerning barred galaxies, see
Sellwood (this volume).

\section{Shapes of rotation curves}

There have been many attempts to infer something about the amount
of dark matter in spiral galaxies by looking at the shape of the 
observed rotation curves. Part of this quest is motivated by the 
desire to arrive at a simple rule of thumb concerning the behaviour
of rotation curves as function of Hubble type and luminosity.
However, as we shall see, the details in individual galaxies are
important at some level, and no such thing as a universal rotation
curve exists.

\subsection{``Wiggles''}

One argument for the maximum disc hypothesis is the presence of
``wiggles'' in the rotation curves, in particular those derived
from long slit spectroscopy. This was already remarked upon by
Kent (1986) and Van Albada \& Sancisi (1986). These wiggles can,
remarkably, be fitted using the maximim disc approach, even though
it is quite likely that the ``wiggles'' in the position - velocity
diagram resulting directly from the spectrocopy are due to the 
crossing of the spiral arms, as in the case of NGC 2998 (cf. Rubin 
et al. 1978). Freeman (1992) likewise shows a few cases from the 
work of Buchhorn (1992) using the Mathewson, Ford \& Buchhorn (1992) data.

However, when the motions giving rise to the ``wiggles'' are non-circular
motions due to the spiral arms, it is technically incorrect to speak of
rotation curves here. Visser (1980), in his study of M81, shows that the
average rotation curve of a two-dimensional velocity field strongly
perturbed by motions due to the spiral arms is wiggly, while the ``true''
rotation curve he needs to fit such a velocity field with a model based
on density wave streaming motions is smooth and without ``wiggles''.
Thus the ``wiggle'' argument is not usable in the context of rotation
curves. In any case, Van der Kruit (1995) shows that also for non-maximum
discs the wiggles can be fitted.

The only rational way to reformulate the ``wiggle'' argument, as
proposed by Binney (priv. comm.), is to consider that, since they are 
there and due to spiral arm streaming motions, there should be a 
non-negligible part of the mass in the disc. A similar argument is 
made by Sellwood (this volume) concerning the capability of fitting 
gas flow models of barred spirals to high resolution two dimensional 
velocity fields.

\subsection{Complete rotation curves}

A first requirement to discuss rotation curve shapes is to have
them defined over as large a range in radius as possible. This means
that HI rotation curves need usually to be supplemented with optical
data in the inner parts, to assure a better definition of the curve 
there. This can be done by e.g. long slit spectra of optical emission
lines - usually H$\alpha$ and [NII] - or, preferably, with 
two-dimensional Fabry-P\'erot data. Alternatively, CO observations 
in the millimeter wavelength range can be used, either from single 
dish data, or from interferometry. Sofue (1997) published a number of 
combined rotation curves using available data in HI, CO and H$\alpha$.

Begeman (1989) developed a method to correct HI rotation curves
for the effects of beam smoothing, using high sensitivity data for 
the galaxy NGC 3198. He fits multiple gaussians to the line profiles,
and retains the principal peak. His results have been compared with
long slit optical emission line data by Bottema (1988) and Hunter
et al. (1986), and with Fabry-P\'erot data by Corradi et al. (1991).
The agreement between all these data is excellent.

The necessity of good data in the inner parts cannot be overstressed,
since in particular for the multi-component mass models, the outcome
of the decomposition depends on a correct determination of the
rotation curve at all radii. This is particularly important for dwarf 
galaxies, where the disc rotation curve is often compared with 
observational data consisting of only a couple of data points.

\subsection{Conspiracies ?}

In the late seventies, the emphasis on the rotation curve shapes
concerned their flatness~: the absence of the expected Keplerian decline 
of the rotation curve beyond the optical image implied the presence of
a large amount of mass with high mass-to-light ratios (Bosma 1978). 
The systematics of the rotation curves as function of Hubble type 
and luminosity were determined by Rubin and her collaborators in a 
series of papers (Rubin et al. 1978, 1980, 1982, 1985).

In the mid-eighties arguments centered on the flat and featureless
nature of the curves (Bahcall \& Casertano 1985). The question how
a thin disc and a presumably round dark halo conspire to keep the
rotation curve flat without showing any marked feature was raised,
and an answer was found in the process of adiabatic compression
of the halo material when the disc was formed (cf. Barnes 1987,
Blumenthal et al. 1986).

However, some galaxies have rotation curves which decline just beyond 
the optical image, and stay more or less flat thereafter. Early examples
are NGC 5033 and NGC 5055 (Bosma 1978, 1981a), and also NGC 5908
(Van Moorsel 1982). Thus the notion of featureless rotation curves 
never corresponded to hard reality. Two more examples, NGC 2683 and NGC 
3521, were given by Casertano \& Van Gorkom (1991), who speculated that 
declining curves are linked with discs having short scalelengths. However,
Broeils (1992) finds cases of declining curves for galaxies with large
disc scalelengths. 

In any case, all these examples show that the process of adiabatic 
compression is not fully operational for all spiral galaxies. In
particular, a partial decoupling between disc and halo is seen in a 
non-negligible fraction of the brighter spirals.

\subsection{Universal Rotation Curves ?}

Persic, Salucci \& Stel (1996) propose that the shapes of rotation
curves follow a systematic pattern as function of luminosity only.
Their conclusions are based on a re-analysis of the data by Mathewson,
Ford and Buchhorn (1992), although the notion of universal rotation
curve is older (e.g. Rubin et al. 1980). The rotation curves for low 
luminosity galaxies are rising, those of intermediate luminosity are 
more or less flat, while those of the highest luminosity galaxies are 
falling.

However, a cursory inspection of the 12 curves with the highest rotation
velocities in the sample of Persic et al. (1996, see also Persic \&
Salucci 1995 for more details) shows that their weighted average is
not declining at all. Moreover, about 18\% of the galaxies in their 
sample have inclinations larger than 85$^{\circ}$, and about half of 
these are seen exactly edge-on. This has consequences for the shape of
the resulting average rotation curves : as has been shown by Bosma et 
al. (1992) and Bosma (1995), the opacity in the inner parts of large 
spirals tends to make the inner slopes of H$\alpha$ rotation curves 
too shallow compared to the true rotation curves. Thus the averaging 
procedure should be redone without taking the data from highly
inclined galaxies into account.

Even so, Verheyen (1997) reexamined the problem of universal curves, and
finds that for the 30 rotation curves in his sample of spiral galaxies
in the Ursa Major cluster about 10 do not fit at all to the universal
rotation curves of Persic et al. (1996). This means that although there
is some merit to the scheme, at the 10\% - 20\% level the notion of 
universal rotation curves breaks down.

\section{Spiral structure constraints}

\subsection{Swing amplifier criteria}

Athanassoula et al. (1987) introduce criteria from spiral
structure theory, and in particular those for swing amplification 
(Toomre 1981), in order to get limits on the dynamical importance of 
the disc. Since observed discs are usually bisymmetric, it seems
natural to ask for the possibility to have amplification of m = 2 
structures, which imply a minimum for (M/L)$_{\sf disc}$. On the
other hand, one would like to suppress m = 1 structures, which
implies an upper limit for (M/L)$_{\sf disc}$. This limit usually
co\"incides with the requirement to have ``maximum discs'' but with
halos with non hollow cores. Athanassoula et al. (1987) find that
these latter models are preferred when considerations of stellar 
populations and the buildup of Sc discs at a constant rate of star 
formation over a Hubble time are taken into account.

\begin{figure}
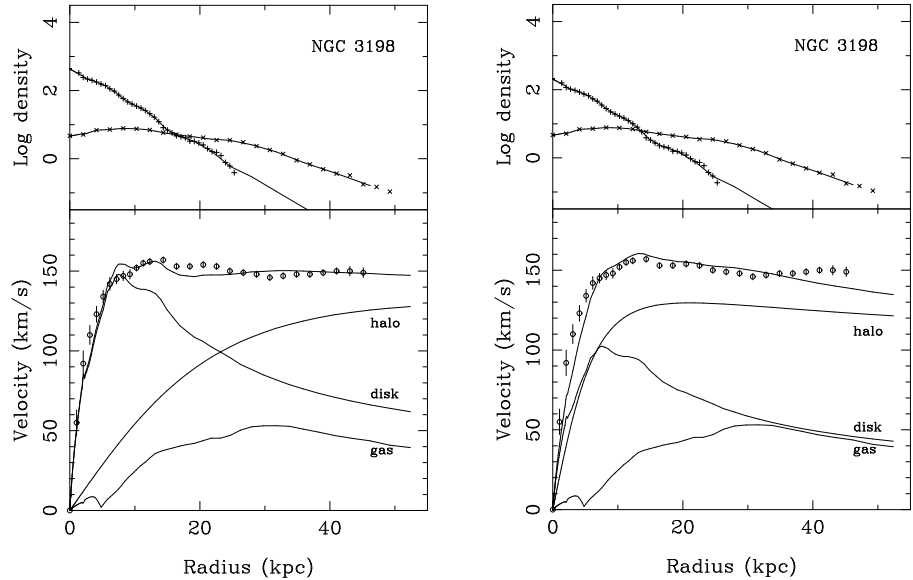

\plotfiddle{figure1a.ps}{0.0truecm}{0}{60}{60}{-365}{-455}
\plotfiddle{figure1b.ps}{7.0truecm}{0}{60}{60}{-183}{-243}
\caption{Mass models for NGC 3198. At left a ``maximum disc'' model,
at right a ``no m = 2'' model. Data from Wevers et al. (1986) 
for the photometry, and Begeman (1989) for the rotation curve.
A Hubble constant of 50 km s$^{-1}$ Mpc$^{-1}$ is assumed.}
\end{figure}

The way this works is illustrated in Figures 1 and 2 for the galaxy 
NGC 3198. For the swing amplifier to work, we need several quantities
which depend on the rotation curve. These are the parameters $\Gamma$
and X, given by :

\begin{equation}
\Gamma = - {R \over \Omega} {d\Omega \over dR} 
\end{equation}

\noindent
where $\Gamma$ is the dimensionless shear rate. This quantity is 1.0
for exactly flat rotation curves, 1.5 for Keplerian curves and 0.5 for
a curve rising as the square root of the radius. The other quantity is

\begin{equation}
X = {{\kappa^2R} \over {2{\pi}Gm{\mu}}}
\end{equation}

\noindent
As can be seen, the active disc mass, $\mu$, comes in, as well as
the number of arms, m. The rotation curve is also represented via the 
epicyclic frequency, $\kappa$.

Athanassoula (1984) rediscussed the swing mechanism presented by
Toomre (1981), and calculated for various values of $\Gamma$ the
maximum growth factor of the swing amplification as function of
X for 3 typical values of the Toomre parameter Q. 
In Athanassoula et al. (1987) we use an interpolation method
to determine the amplification factor for any value of $\Gamma$ and X.
As a result, we can for a given mass model work out its consequences
for the amplification of m = 1, 2, 3, ... structures, and calculate
graphs such as presented in Figure 2. It can be easily seen from
Figure 2 that if we lower the mass-to-light ratio of the disc with
a factor 2, the curves for m = 2 become those in the top panel, and
the curves for m = 4 those in the middle panel (since m$\mu$ is
what matters).

\begin{figure}
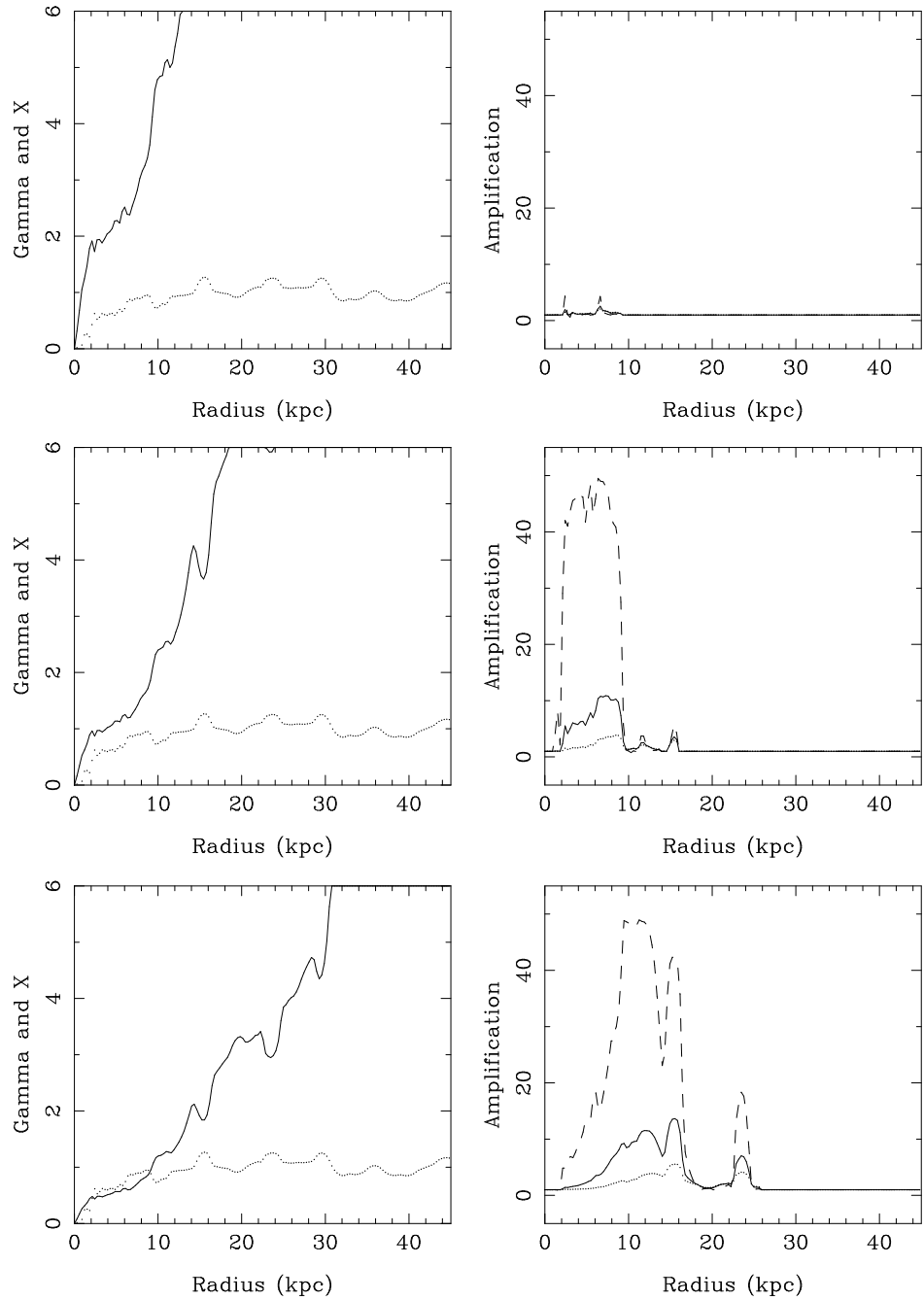

\plotfiddle{figure2a.ps}{0.0truecm}{0}{65}{65}{-210}{-495}
\plotfiddle{figure2b.ps}{0.0truecm}{0}{65}{65}{-210}{-640}
\plotfiddle{figure2c.ps}{16.3truecm}{0}{65}{65}{-210}{-335}
\caption{Curves of $\Gamma$ (dotted line) and X (continous line), 
at left, and the maximum
amplification for three values of Q (Q = 1.2 : dashed line, Q = 1.5
: continuous line and Q = 2.0 : dotted line), at right, for the 
``maximum disc'' model of NGC 3198 shown in Figure 1. At the top, 
the results for m = 1, in the middle, the results for m = 2, and 
at the bottom, the results for m = 4.}
\end{figure}

For NGC 3198, it is clear that the ``no m = 2'' criterion leads to
a disc rotation curve whose maximum velocity is 105 km s$^{-1}$.
This can be compared to the value of 100 $\pm$ 19 km s$^{-1}$ derived by 
Bottema (1993) from his criterion based on his velocity dispersion
work, and also by the maximum values of 93 and 100 km s$^{-1}$ 
calculated by Navarro (1998) for his models, which are partly based 
on notions from cosmological simulations, and include tacitly that 
the adiabatic compression of the halo material due to the disc
formation is functioning as predicted. While the Bottema
criterion is still in rough agreement with the swing amplifier
criteria given the uncertainties, the models of Navarro (1998) are 
not anymore : for them the disc is just a bit too light to allow swing
amplification of m = 2 structures.

In conclusion, the method used by Athanassoula et al. (1987) gives
a {\it {range}} of values for the mass-to-light ratio of the
disc, with the upper limit very often close to or identical with
a ``maximum disc with no hollow core halo'' solution, and a lower
limit which is higher than the values current fashionable arguments 
from cosmological simulations seem to advocate.

\subsection{Recent applications of swing amplifier criteria}

Vogt et al. (1996) began a campaign to obtain an idea of the rotation 
curves of faraway galaxies, thanks to powerful equipment like the Hubble 
Space Telescope and the Keck Telescope. For about a dozen spirals they 
derive position velocity curves, which are heavily affected by aperture
effects, since the slit width is non-negligible compared to the size of
the galaxies. For one galaxy, 0305-00115, at z = 0.48, Fuchs,
M\"ollenhoff \& Heidt (1998) argue on the basis of swing amplifier 
criteria that a dark halo ought to be present~: for a maximum disc 
solution the number of arms, calculated on the assumption that X = 2, 
is too low. However, in the rising parts of the rotation curve
$\Gamma$ is less than 1.0, so that the X-value corresponding to 
maximum amplification is less than 2.

Mihos, McGaugh \& De Blok (1997) apply  the same notions of stability
criteria to the Low Surface Brightness (LSB) galaxies. Since these are
thought to be dark matter dominated, the amplification of m = 2 structures
in those galaxies is much more difficult. Only interactions can trigger
m = 2 disturbances, and even then the amplitudes are small and the
deviations for axisymmetry not very large. However, the decomposition
into disc and halo depends crucially on the correct determination of
the rotation curve in the inner parts, and, as shown by De Blok (this
meeting), the data for the LSB galaxy UGC 128 show that its disc is
not necessarily dark matter dominated in the inner parts.

\begin{figure}
\plotfiddle{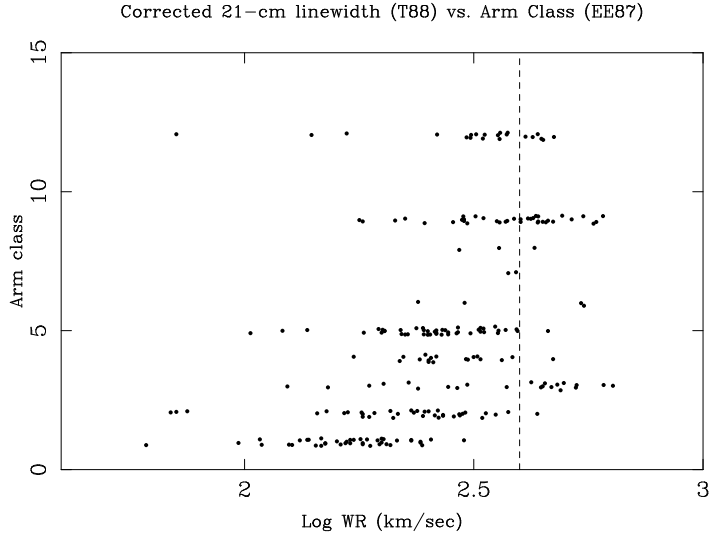}{7.0truecm}{-90}{40}{40}{-180}{230}
\caption{HI line width corrected for inclination according to the
data given in Tully (1988) as function of arm class for the galaxies
in the compilation of Elmegreen \& Elmegreen (1987). The dashed line 
corresponds to a corrected line width of 400 km s$^{-1}$}.
\end{figure}

\subsection{Flocculent and grand design galaxies}

Elmegreen \& Elmegreen (1990) made a statistic based on an extended
gradient in the rotation curve, and compared this with their arm 
classes (Elmegreen \& Elmegreen 1982). They concluded that there is
a correlation between the two, in the sense that declining rotation curves
are found in grand design spirals. From this they concluded that grand
design spirals have smaller dark halos than flocculent galaxies.

Their interpretation, however, rests on their mixing large and small
galaxies into only three bins, flocculent galaxies, intermediate galaxies,
and grand design galaxies. A more detailed analysis can be done if one 
considers the width of the HI profile, corrected for inclination, with 
arm class. This has been done in Figure 3 for all the galaxies for
which arm classes are available in Elmegreen \& Elmegreen (1987) and
HI profile data in Tully's (1988) {\it Nearby Galaxy Catalog}. 
It can be seen that the galaxies in arm class 3 have high rotational 
velocitites, which corresponds with their regular structure. Combining
arm classes 1 - 4 into one bin called ``flocculent'' thus lumps
together dwarf galaxies with rising curves and giant galaxies with 
slowly declining curves. The ``grand design'' galaxies (arm classes 
9 \& 12) are predominantly giant galaxies, and the intermediate
classes (arm classes 5 - 8) also. In view of these biases,
Elmegreen \& Elmegreen's (1990) conclusions cannot be sustained.

\subsection{Declining rotation curves}

Rotation curves which drop relatively sharply beyond the optical
radius and stay more or less flat thereafter, might, because of the 
additional identifiable feature, hold out a promise to enable
us to discriminate between the various mass models. Since one
expects them to be found amongst galaxies with high rotational 
velocities, I made a small survey with the VLA of a number of galaxies
with W$_{\rm R}$ $>$ 400 km/s in collaboration with Van Gorkom, Gunn,
Knapp and Athanassoula. Several new cases of galaxies with such declining
rotation curves were found. In Bosma (1998) a preliminary account is 
given for the most spectacular case, NGC 4414, for which also radial 
velocities and velocity dispersion information was obtained.

Unfortunately, the range in disc mass-to-light ratios for that galaxy
cannot be constrained very easily, in spite of the feature. However,
the velocity dispersion data allow the evaluation of the Toomre 
Q-parameter, which is found to be about 1.15 for a maximum disc model,
but 2.30 for a ``no m = 2'' model. The latter value is definitely too
high to allow spiral structure from swing amplification. 
A weak global spiral pattern is present in the old disc (Thornley
1996). Therefore, it seems unlikely that the inner parts of
bright disc galaxies are dark matter dominated.

Another spectacular case of a declining rotation curve has been
reported for NGC 157 by Ryder et al. (1998). The strong warp and
the relatively low inclination of the HI envelope make the
determination of the rotation curve a bit uncertain, but there is a 
definite decoupling between the inner disc and the outer halo as
traced by the HI.

\section{Velocity dispersions}

\subsection{NGC 3198 once more}

Bottema (1993), from an analysis of velocity dispersions, claims
that the maximum velocity of the disc component is 63\% of the
maximum observed velocity. The path to this result is strewn with
assumptions, the most important of which are that discs are
exponential with a velocity ellipsoid close to that in the solar
neighbourhood, that Freeman's (1970) law holds, and that
(B-V)$_{\sf old~disc}$ = 0.7 for all discs. As already discussed above,
for NGC 3198 Bottema's result corresponds closely to the ``no m = 2''
solution proposed by Athanassoula et al. (1987). 

A quick re-evaluation of Bottema's result can be formulated as
follows. From the good agreement between the stellar and gas
rotation data presented in Bottema (1988), it follows that
the asymmetric drift in NGC 3198 is small, and thus the epicyclic 
approximation can be used. This states that $\kappa$/2$\Omega$ = 
$\sigma$$_{\phi}$/$\sigma$$_{R}$, where $\Omega$ is the angular 
velocity, $\kappa$ is the epicyclic frequency, and $\sigma$$_{R}$ 
and $\sigma$$_{\phi}$ are respectively the radial and azimuthal
velocity dispersion. From the measurement of the
rotation curve and the velocity dispersion along the major axis,
ignoring the z-axis contribution for the moment, we can then calculate
the radial velocity dispersion $\sigma$$_{R}$ as function of radius,
and from the mass model we can calculate the critical radial velocity
dispersion needed for axisymmetric stability. The ratio between
these two is Toomre's (1964) parameter Q~:

\begin{equation}
Q = {{\sigma{_{_{R}}}}\over{\sigma{_{crit}}}} = 
{{\sigma{_{_{R}}}\kappa}\over {3.36G\mu}}
\end{equation}

\noindent
For NGC 3198 we then find for the ``maximum disc'' model a mean
Q of 0.92, and for a ``no m = 2'' model a mean Q of 1.92. The real
uncertainties in these values are about 20\%. The important result
is that indeed for NGC 3198 the measured velocity dispersion seems
too low to have a stable ``maximum disc'' solution (see also Fuchs's
contribution in this volume). 

One can argue with this in several ways. First of all, the
observed major axis velocity dispersion is a combination of
$\sigma$$_{\phi}$ and $\sigma$$_{z}$, so the determination of
$\sigma$$_{R}$ may not be entirely correct. 

Second, the numerical 
factor 3.36, which corresponds to a Schwarzschild distribution of 
the random velocities in an infinitesimally thin disc, should perhaps be 
corrected for the effect of thickness, possibly also for a different 
shape of the distribution function, and certainly for the effect of 
the gas which is important for a late type spiral. The thickness 
correction alone (calculated assuming all the material in the disc,
and the solar neighbourhood value for the ratio of vertical to
radial velocity dispersions) lowers the numerical factor to 2.6 
(cf. Shu 1968, Vandervoort 1970) and the effect of 10\% of gas 
enhances then the factor to about 2.9 (cf. Toomre 1974). Other
distribution functions may apply~: Fuchs \& Von Linden (1998) 
rediscuss and extend work by Graham (1967) and by Toomre (unpublished)
on an exponential distribution function, for which the
numerical factor is 3.944 instead of 3.36, so that with the 
thickness and 10\% gas corrections we end up at 3.40. From
Graham's (1967) Figure 1, where he shows results for several
other distribution functions, it becomes clear that the uncertainty
in the numerical factor could be easily 20 - 30\%.

Finally, as
argued by Kormendy (priv. comm., see also discussion after Fuchs's
contribution), the influence of younger stellar populations in the 
spectra could result in lower measured velocity dispersions. In 
view of all these arguments, perhaps we should not be overly 
worried about the result that Q is just below 1.00 in NGC 3198.

\subsection{Other galaxies}

As discussed already above, for NGC 4414 I find a mean Q of about
1.15 for a ``maximum disc'' model, and about 2.30 for a ``no m = 2'' 
model. For NGC 2841, a preliminary analysis shows that for a ``maximum
disc'' model Q $\simeq$ 1.5 - 1.6, so that here a ``no m = 2'' model is
dynamically too hot to be acceptable. Further work on other galaxies
is in progress. In all cases, spectra are taken along the minor
and the major axes, since only their combination can help us
disentangle the radial, azimuthal and perpendicular velocity dispersions
(cf. work on NGC 488 by Gerssen et al. 1997). It is hoped that
with such data, a clearer answer can be given to the question
whether the disc is close to ``maximum disc'' or not.

\section{Halo parameters}

Once halo parameters are derived, what can one do with them ?
Scaling relations between the halo parameters have been reviewed by 
Kormendy (1988), partly on the basis of the ``no m = 1'' models and
results of Athanassoula et al. (1987). Apart from
relatively obvious correlations, e.g. between V$_{\sf max, disc}$ and
the velocity dispersion of the halo, there are a few which are
intruiging, such as the (weak) relation between the ratio of halo 
core radius and optical radius with Hubble type, and the correlation
between halo core radius and central density of the halo for Sc -
dwarf galaxies. These relations will be further investigated (Kormendy,
priv. comm.).

Bosma (1991) shows that the frequency of warps, which is at least
50\% for all spirals, depends on the
ratio of halo core radius to optical radius of the galaxy : galaxies
for which this ratio is small do not have warped HI discs. This is
usually attributed to dynamical friction between a misaligned disc
and a dark halo. If the dark halo is strongly concentrated, such
misalignments are short lived, as is shown also by numerical
simulations (Dubinski \& Kuijken 1995).

Finally, Navarro (1998) uses the information on a concentration
index, derived by fitting a functional form to the observed 
rotation curves, to argue that dwarf galaxies are too concentrated
for a standard CDM $\Omega$ = 1 model, but favour instead cosmological
models with $\Omega <$ 1.

\acknowledgments

I would like to thank Lia Athanassoula for fruitful discussions and
comments on a draft of this paper. I gratefully acknowledge
discussions with James Binney, Burkhard Fuchs, John Kormendy, Jerry 
Sellwood and Peter Vandervoort concerning some of the topics raised
in this review.

\end{document}